\documentclass[11pt]{article}
\usepackage{graphicx}
\usepackage{rotating}

\newcommand{\BABARPubYear}    {01}

\newcommand{\BABARConfNumber} {24}
\newcommand{\SLACPubNumber} {8910}

\input pubboard/babarsym

\long\def\inst#1{\par\nobreak\kern 4pt\nobreak
    {\it #1}\par\vskip 10pt plus 3pt minus 3pt}

\begin{document}
{\pagestyle{empty}

\begin{flushright}
%BaBar Analysis Document \#240, Version 4\\
\babar-CONF-\BABARPubYear/\BABARConfNumber \\
%\babar-PUB-\BABARPubYear/\BABARPubNumber \\
SLAC-PUB-\SLACPubNumber \\
%hep-ex/\LANLNumber \\
July 10, 2001 \\
\end{flushright}

\par\vskip 5cm

% Title of the paper
\begin{center}
\Large \bf\boldmath Search for the Rare Decays 
$B \rightarrow K \ell^+ \ell^-$ and  $ B \rightarrow K^{\ast}(892) \ell^+ \ell^-$  
\end{center}
\bigskip

\begin{center}
\large The \babar\ Collaboration\\
%\mbox{ }\\
%\today
\end{center}
\bigskip \bigskip

% Abstract
\begin{center}
\large \bf Abstract
\end{center}
We present preliminary results from a search for the flavor-changing neutral current
decays $B\to K\ell^+\ell^-$ and $B\to K^*(892)\ell^+\ell^-$ using a sample of 
$22.7\times 10^6$
$\Upsilon(4S)\to B\bar B$ decays collected with the \babar\ detector at the \pep2 \BF. We have
reconstructed the following final states: $B^+\to K^+\ell^+\ell^-$, 
$B^0\to K^0\ell^+\ell^-$ ($K_s^0\to\pi^+\pi^-$),
$B^+\to K^{*+}\ell^+\ell^-$ ($K^{*+}\to K_s^0\pi^+$),
and $B^0\to K^{*0}\ell^+\ell^-$ ($K^{*0}\to K^+\pi^-$),
where $\ell^+\ell^-$ is either an $e^+e^-$ or $\mu^+\mu^-$ pair. We obtain 
the 90\% C.L.~upper limits  
${\mathcal B}(B\to K\ell^+\ell^-)< 0.6\times 10^{-6}$ and 
${\mathcal B}(B\to K^*\ell^+\ell^-)<2.5\times 10^{-6}$, close to
the Standard Model predictions for these branching fractions.
\vfill
\begin{center}
Submitted to the International Europhysics 
Conference on High Energy Physics, \\
7/12---7/18/2001, Budapest, Hungary
\end{center}

\vspace{1.0cm}
\begin{center}
{\em Stanford Linear Accelerator Center, Stanford University, 
Stanford, CA 94309} \\ \vspace{0.1cm}\hrule\vspace{0.1cm}
Work supported in part by Department of Energy contract DE-AC03-76SF00515.
\end{center}

\newpage
}
% Input author list file
\begin{center}
\small

The \babar\ Collaboration,
\bigskip

%% author list as of 12-Jul-2001 (622 authors)
B.~Aubert,
D.~Boutigny,
J.-M.~Gaillard,
A.~Hicheur,
%A.~Jeremie, per J.P.Lees
Y.~Karyotakis,
J.~P.~Lees,
P.~Robbe,
V.~Tisserand
\inst{Laboratoire de Physique des Particules, F-74941 Annecy-le-Vieux, France }
A.~Palano
\inst{Universit\`a di Bari, Dipartimento di Fisica and INFN, I-70126 Bari, Italy }
G.~P.~Chen,
J.~C.~Chen,
N.~D.~Qi,
G.~Rong,
P.~Wang,
Y.~S.~Zhu
\inst{Institute of High Energy Physics, Beijing 100039, China }
% ...

\end{center}\newpage

%%%%%%%%%%%%%%%%%%%%%%%%%%%%%%%%%%%%%%%%%%%%%%%%%%%%%%%%%%%%%%%%%%%%%%%%
% The body of the paper starts here

\section{Introduction}

The flavor-changing neutral current (FCNC) decays 
$B\to K\ell^+\ell^-$ and $B\to K^*(892)\ell^+\ell^-$, 
where $\ell^{\pm}$ is 
a charged lepton, are highly suppressed in the Standard 
Model, with branching fractions predicted to be of order 
$10^{-7}-10^{-6}$. The dominant contributions arise at
the one-loop level and involve a $b\to t\to s$ 
transition, so these decays are sensitive to
the Cabibbo-Kobayashi-Maskawa factor $V_{ts}^*V_{tb}$. 
The loop can involve either the
emission and reabsorption of a virtual $W$-boson,
with the radiation of a virtual photon or $Z$ (that subsequently 
materializes into the $\ell^+\ell^-$ pair), or the emission
of two virtual $W$ bosons, producing the $\ell^+\ell^-$ pair
through a box diagram. Such processes are known as
electroweak penguins.
The simpler decays $B\to K^*\gamma$
and $B\to X_s\gamma$ (where $X_s$ is any hadronic system with 
strangeness) have been observed~\cite{bib:CLEOKstargam,bib:CLEOXsgam},
providing the first evidence for the electromagnetic 
penguin amplitude.

These rare decays are interesting not only as a probe of
Standard Model loop effects, but also because their rates
and kinematic distributions are sensitive
to new, heavy particles that can appear 
virtually in the loop~\cite{bib:AliBallPRD}. 
Such heavy particles are predicted, for 
example, by supersymmetry (SUSY) models.
 
Table~\ref{tab:predictions} lists the predictions of a number of calculations
based on the Standard Model. In such calculations, the short-distance
physics that governs free-quark decay 
is incorporated into the Wilson coefficients 
$C_{7}^{\rm eff}$, $C_9^{\rm eff}$, and $C_{10}$ in
the Operator Product Expansion~\cite{bib:Buras} of the effective Hamiltonian.
This part of the calculation is relatively well 
understood in the Standard Model, except for nonperturbative
contributions in the dilepton mass regions near the charmonium resonances.
% The Wilson coefficient $C_9^{\rm eff}$ is a function of $m_{\ell\ell}$;
% it receives long-distance physics contributions when $m_{\ell\ell}$ is
% at or near the mass of one of the charmonium resonances.
There are, however, significant uncertainties in the predicted rates,
which arise from strong (QCD) interactions among the quarks.
Although these long-distance effects are difficult to calculate, they
can be rigorously parametrized in terms of form factors,
which are estimated with methods such as light-cone QCD sum rules
or lattice QCD. From Table~\ref{tab:predictions}, we see that the
rate for $B\to K^*\ell^+\ell^-$ is expected to be three to four times
larger than that for $B\to K\ell^+\ell^-$. The amplitudes for the 
$B\to K^*\ell^+\ell^-$ modes include a pole at 
$q^2=m^2_{\ell^+\ell^-}=0$, 
where the photon
is on the mass shell. This pole gives a substantial contribution 
to the rate for kinematic configurations
in which $q^2\approx 0$ and is particularly
large in the $B\to K^* e^+e^-$ mode. Measurements of $B\to X_s\gamma$
constrain the magnitude of $C_7^{\rm eff}$, the coefficient of the
electromagnetic penguin operator, but new physics could 
still affect the phase of $C_7^{\rm eff}$.

\begin{table}[b]
  \begin{center}
    \begin{tabular}{lccc}
    \hline\hline
                            &  \multicolumn{3}{c} {$\mathcal{B}$/10$^{-6}$}   \\  \cline{2-4}
   \multicolumn{1}{c}{Model}  & $K \ell^+ \ell^-$ & $K^{\ast} e^+ e^-$ & $K^{\ast} \mu^+ \mu^-$\\
     \hline
 LCSR~\cite{bib:AliBallPRD}          &  $0.57^{+0.17}_{-0.10}$ & $2.3^{+0.7}_{-0.5}$ & $1.9^{+0.5}_{-0.4}$ \\

 LCSR~\cite{bib:Aliev1997,bib:Aliev2000}            &            &                 & $1.4$  \\

 Quark Models~\cite{bib:Melikhov1998,bib:Melikhov2000}      
                                     &  $0.62\pm0.13$ & $2.1\pm0.7$ & $1.5\pm0.6$ \\

% Lattice QCD~\cite{bib:DelDeb1998}   &   & 1.15 & 1.15 \\

 QCD Sum Rules~\cite{bib:Colangelo1996,bib:Colangelo1999} 
                                     &  0.3 &  & 1.0 \\
    \hline \hline
    \end{tabular}
  \end{center}
  \caption[Branching fraction predictions in the Standard Model]
    {\label{tab:predictions}
       Branching fraction predictions in various models within the Standard
Model framework: light cone QCD sum rules (LCSR), quark models, and QCD sum rules.
   }
\end{table}                 

We search for $B$-meson decays in the following channels:
$B^+\to K^+\ell^+\ell^-$, 
$B^0\to K^0\ell^+\ell^-$ ($K_s^0\to\pi^+\pi^-$),
$B^+\to K^{*+}\ell^+\ell^-$ ($K^{*+}\to K_s^0\pi^+$),
$B^0\to K^{*0}\ell^+\ell^-$ ($K^{*0}\to K^+\pi^-$)
where $\ell^+\ell^-$ is either an $e^+e^-$ or a $\mu^+\mu^-$ pair.
Throughout this paper, charge conjugate modes are implied.

%  Detector description

\section{Detector Description and Data Samples}

The data used in the analysis were collected with the \babar\ detector
at the \pep2\ storage ring at the Stanford Linear Accelerator Center.
We analyzed the data taken in the 1999--2000 run, consisting
of a 20.7 \invfb\ sample taken on the $\Upsilon(4S)$ resonance, as well as 
2.6 \invfb\ taken at
a center-of-mass energy 40 MeV below the $\Upsilon(4S)$ resonance peak 
to obtain a pure continuum sample. 
Continuum events include non-resonant
$e^+e^-\to q\bar q$ production, where $q=u$, $d$, $s$, or $c$.                 
The on-resonance sample contains
$(22.7\pm0.4)\times 10^6$ $\Upsilon(4S)\to B\bar B$ events.

The \babar\ detector is described in detail elsewhere~\cite{bib:babarNIM}. 
All components of the detector were used for this study. Of particular importance for this
analysis are the charged-particle tracking system and the detectors used for
particle identification. At radii between about 3 cm and 14 cm,
charged tracks are measured with high precision in a five-layer
silicon vertex tracker (SVT).  Tracking beyond
the SVT is provided by the 40 layer drift chamber, which extends in radius 
from 23.6 cm to 80.9 cm.
Just outside the drift chamber is the DIRC, which
is a Cherenkov ring-imaging particle identification system. Cherenkov light
is produced by charged tracks as they pass through an array of 144
five-meter-long fused silica quartz bars. The Cherenkov light is transmitted to
the instrumented end of the bars by total internal reflection, preserving the
information on the angle of the light emission with respect to the
track direction. The DIRC is used for kaon identification in this
analysis and is essential to our background rejection.
Electrons are identified using
an electromagnetic calorimeter comprising 6580 thallium-doped CsI
crystals. Muons are identified in the Instrumented Flux Return (IFR),
in which resistive plate chambers (RPCs) are interleaved
with the iron plates of the flux return.

%  a little comment from Jeff to test CVS
%  Data and event selection

\section{Overview of Analysis Method}

Our search for $B\to K^{(*)}\ell^+\ell^-$ was guided by several
key considerations. Given the predicted branching fractions
and our reconstruction efficiencies (6\% to 18\%), 
we expect only a handful of observed signal 
events. Because the experimental signature is strong, however, it
might be possible to establish a signal with a small number of
events, as long as effective background suppression is
achieved. But in working with small event samples,
we must ensure that no bias is introduced through the event selection that 
could produce either an artificial enhancement of events in the signal region,
or an artificial suppression of events in the sidebands used to determine
the background level.

To avoid the possibility of bias, both the signal
region and the sidebands used to measure the background levels are
``blinded'' during the process of defining the event selection criteria,
so that statistical fluctuations in these regions cannot
be exploited to artificially enhance a signal. Once the event
selection is defined, we apply the cuts, unblind the data, 
and perform unbinned 
maximum likelihood fits to determine the signal and background 
yields. In these fits, the overall background level is determined 
from the data, with shapes derived from Monte Carlo but checked
against control samples from the data.
Although we define signal and sideband regions for the purpose of
optimizing event selection cuts, the distinction between these regions
is not relevant to the fit. 

To optimize the event selection without using either the signal region
or the sidebands used for the final background determination,
we have used
(1) full \geant3~\cite{bib:GEANT} Monte Carlo for signal and backgrounds,
(2) high statistics fast Monte Carlo for backgrounds, and
(3) a continuum data sample collected at a center-of-mass
energy below the $\Upsilon(4S)$.
For simulation of signal events, we have
used the event generator \evtgen~\cite{bib:evtgen} to implement a full
form-factor parametrization of the decay amplitudes. The
simulation provides the detailed kinematic correlations
among the particles; in particular, the joint distribution
in the Dalitz plot variables $q^2$ and lepton energy is
simulated. Final-state photon radiation is simulated with
the {\tt PHOTOS}~\cite{bib:PHOTOS} package.
The selection procedure was optimized using matrix elements
from the model by Ali {\it et al.}~\cite{bib:AliBallPRD}, but 
the predictions from a number of other models were studied
as well. About twenty thousand events were 
generated for each signal final state using a full
\geant3 simulation. 

To validate the Monte Carlo modeling of both the signal and
backgrounds, and to determine some of the systematic errors, 
we have used an extensive set of control samples from the data.
The modeling of distributions characterizing 
signal events, and the associated cut efficiencies,
were checked using $B\to J/\psi K$ and $B\to J/\psi K^*$ decays 
(with $J/\psi\to \ell^+\ell^-$). These decays yield final states that
are topologically identical
to signal events (although they have a very narrow distribution 
in $q^2=m^2_{\ell^+\ell^-}$). These modes are both serious backgrounds that
must be suppressed and powerful control samples for understanding the signal efficiency.
Another control sample is provided by 
reconstructed $Ke\mu$ and $K^*e\mu$ combinations, which cannot result from $B$ decays
in the Standard Model. The Monte Carlo predictions for the background shapes 
associated with combinatoric backgrounds can be checked using these samples.
Certain individual backgrounds, especially 
those that peak under or near the signal, are studied with 
\geant3 Monte Carlo samples of these decays. Finally, we defined a ``large
sideband'' region in the data that is sufficiently far from the signal that we do
not need it for determining the background level, but which allows us to check that 
expectations for background levels based on Monte Carlo are essentially confirmed in
data prior to unblinding.

The major sources of backgrounds are
$\Upsilon(4S)\to B\bar B$ 
and continuum events. More specifically, these include  
(1) $B\to J/\psi K^{(*)}$ or $B\to \psi(2S)K^{(*)}$ events;
(2) generic $B\bar B$ backgrounds with either two real leptons 
or one real lepton and one hadron misidentified as a lepton (usually a muon);
(3) background from continuum processes, especially $c\bar c$ events with a pair of
$D\to K^{(*)}\ell^+\nu$ decays, or events with hadrons faking leptons;
(4) very small but potentially serious contributions from 
a number of $B$ decay modes with similar
topology to the signal, such as $B^+\to D^0\pi^+$ with $D^0\to K^-\pi^+$,
in which hadrons misidentified as leptons
can create a false peaking signal; and (5) $B\bar B$ or continuum events
with photon conversions in the detector material.        

A large number of Monte Carlo samples were generated to study
these backgrounds. We studied
$B\to J/\psi K^{(*)}$ and $B\to \psi(2S)K^{(*)}$ events
using about twenty thousand \geant3 events in
each final state. 
Approximately $8.4$ M \geant3
generic $B\bar B$ events and about 22 M continuum events were generated.
In addition, we have   
developed a fast, parametrized Monte Carlo and 
generated 220 M $B\bar B$ and 690 M continuum events, corresponding to
200 fb$^{-1}$ of integrated luminosity.
The realism of this simulation was enhanced by 
the fact that many key quantities
are simulated using look-up tables 
with detailed information on charged-track 
efficiencies, particle ID efficiencies, and particle ID misidentification
rates, all of which were measured in studies of data.

\section{Event Selection}

In the initial step of the event selection,
we require that each event have at least four charged tracks and that
the ratio of the second and zeroth Fox-Wolfram moments~\cite{bib:FoxWolfram}
be less than 0.5.  This requirement provides a first suppression of 
continuum events,  which have a more collimated (``jet-like'') event 
topology than $B\bar B$ events.  We also require that the events contain two 
oppositely charged loosely identified leptons with laboratory-frame 
momenta \mbox{$p > 0.5 \ (1.0) \ {\rm GeV}/c$} for electron (muon) 
candidates.  Electrons are identified primarily on the basis of 
$E/p$, where $E$ is the energy measured in the CsI electromagnetic 
calorimeter. 
Muons are identified mainly
by the number of nuclear interaction lengths penetrated by the charged 
track through the detector.

The events are also required to lie 
within a large, rectangular region in the plane defined 
by two kinematic variables~\cite{bib:babarNIM}:
%the beam-energy substituted mass 
\mbox{$m_{\rm ES} > 5.0$~\gevcc} and \mbox{$|\Delta E| < 0.8$~\gev}.
Because signal events contain two $B$ mesons and no additional particles,
the energy of each $B$ in the center-of-mass (CM) frame of the $\Upsilon(4S)$ 
must be equal to
the $e^+$ or $e^-$ beam energy in the CM frame. We define
\begin{eqnarray}
m_{\rm ES}&=& \sqrt{(E_{\rm beam}^*)^2 - (\sum_i {\bf p}^*_i)^2} \nonumber\\
\Delta E&=& \sum_i\sqrt{m_i^2 + ({\bf p}_i^*)^2}- E_{\rm beam}^*,\nonumber
\end{eqnarray}
where $E_{\rm beam}^*$ is the beam energy in the CM frame,
${\bf p}_i^*$ is the CM momentum of particle $i$ in the
candidate $B$-meson system, and $m_i$ is the mass of particle $i$. 
For signal events, 
the beam-energy-substituted $B$ mass, $m_{\rm ES}$, peaks at $m_B$ with 
a resolution of about 2.5 MeV. The quantity $\Delta E$ is used to determine
whether a candidate system of particles has total energy consistent with
the beam energy in the CM frame.

We next apply a tighter set of 
particle identification requirements on both leptons and hadrons. 
The very tight selection criteria we apply to electrons give an
efficiency of about 88\% assuming well measured charged tracks, with 
a corresponding probability for a pion to fake an electron 
signature of about 0.15\%. 
For our very tight cuts, the typical muon ID efficiency for 
momenta greater than about 1 GeV/$c$ is 
60\%--70\%, with a pion fake probability of about 2\%.
Electrons and positrons are required to pass a conversions veto, 
which suppresses $e^+e^-$ pairs that could have come from photon 
conversions in the detector material. 
For the  \mbox{$B^0\rightarrow K^{\ast0} \ell^+\ell^-$} channels, we reconstruct the 
$K^{\ast0}$ in the \mbox{$K^+ \pi^-$} final state, while the $K^{\ast+}$ in 
\mbox{$B^+\to K^{\ast+}\ell^+\ell^-$} is reconstructed in the \mbox{$K_S^0 \pi^+$}
final state.  The mass of the $K^\ast$ candidate is required to 
be within \mbox{75$~\mevcc$} of the mean $K^\ast(892)$ mass.  
The charged kaon daughters of the 
$K^\ast$ are required to be identified as kaons, and all charged tracks 
not identified as kaons are taken to be pion candidates.
The kaon identification is based on the measurement of the 
Cherenkov angle from the DIRC for \mbox{$p > 0.6$ GeV/$c$}
and on the energy-loss measurements (\dedx) from the silicon vertex tracker and 
the drift chamber for momenta \mbox{$p <  0.6$ GeV/$c$}. 
The typical charged kaon ID efficiency is about 80\% for well reconstructed tracks, 
with a pion fake probability of about 2\%.
$K_S^0$ candidates are reconstructed by
combining two oppositely charged pions and requiring that the mass of the 
candidate be within \mbox{9.3 MeV/$c^2$} of the mean $K_S^0$ mass,
that the $K_S^0$ candidate vertex probability be greater than \mbox{0.1\%}, and 
that the distance between the vertex and the candidate 
$B$ vertex be greater than \mbox{0.1 cm}. For each final state we select 
at most one combination of particles as a $B$ signal
candidate per event.  If multiple candidates occur, we select the 
candidate with the greatest
number of drift chamber and SVT hits on the charged tracks.

\begin{figure}[!tb]
 \begin{center}
   \includegraphics[width=5.2in]{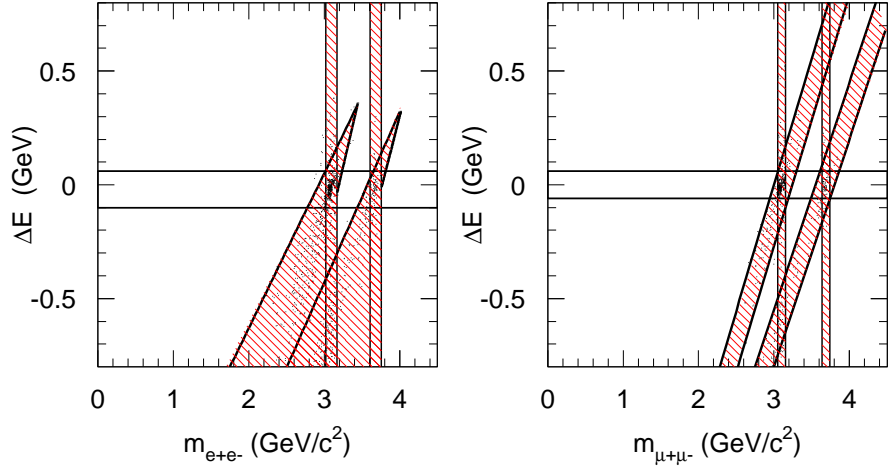}
  \end{center}
  \vspace{-1.2cm}
\caption[Definition of the charmonium veto region.]
{\label{fig:charmoniumveto}
 The veto regions in the $\Delta E$ vs.~$m_{\ell^+\ell^-}$ plane
that are populated by $B\to J/\psi K$ and $B\to \psi(2S) K$ events. 
The shaded areas are vetoed. 
(The same veto is applied to the $K^*$ modes.)
For reference, the
two horizontal lines bound the region in which most signal events are
found. The charmonium vetos remove these backgrounds not only from
the signal region, but also from the sideband region, simplifying
the description of the background in the fits.
}
\end{figure}

Charmonium background, \mbox{$B \to J/\psi K^{(*)}$} and
\mbox{$\psi(2S) K^{(*)}$}, is suppressed by applying a veto in 
the $\Delta E$ vs. $m_{\ell^+\ell^-}$ plane. The boundaries
of the vetos are shown in Fig.~\ref{fig:charmoniumveto}.
It is not sufficient
to remove events with dilepton masses consistent with those of the
$J/\psi$ or $\psi(2S)$, since bremsstrahlung or track 
mismeasurement can result in a large departure of the dilepton
mass from the resonances.  
Charmonium events can pass this veto if one of the leptons and 
the kaon are misidentified as each other (``swapped'').
There is also a fairly significant feed-up from the
\mbox{$B\to J/\psi K$} and \mbox{$B\to \psi(2S) K$} modes to the
\mbox{$B \to K^\ast \ell^+ \ell^-$} channels. Both the
swapped candidates and the feed-up contributions are explicitly 
vetoed.

\begin{figure}[!tb]
 \begin{center}
   \includegraphics[height=7.1in]{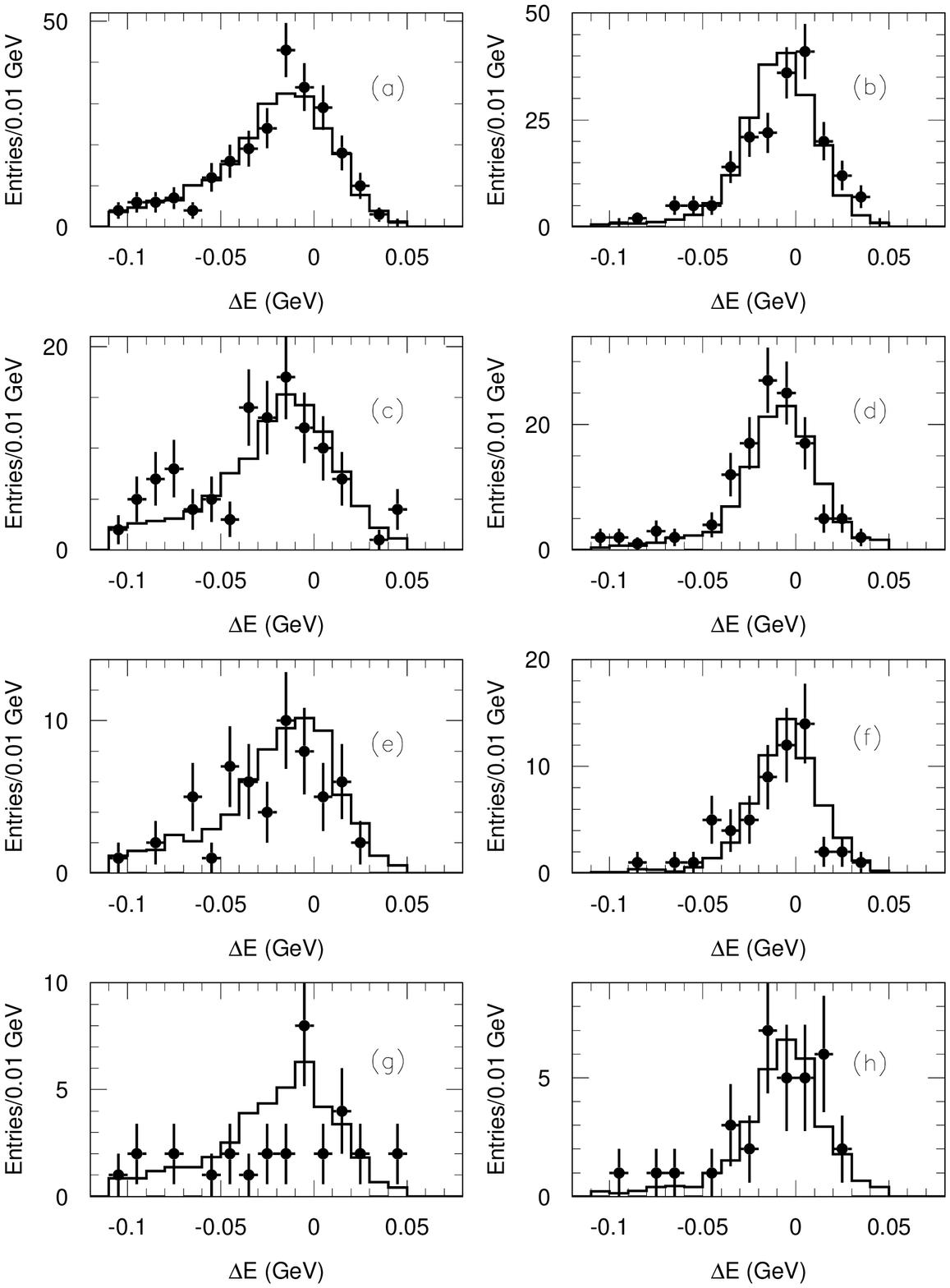}
  \end{center}
  \vspace{-1.2cm}
\caption[Comparison of $\Delta E$ shapes in the charmonium control sample]
{\label{fig:jpside}
 Comparison of $\Delta E$ shapes between data and Monte Carlo for the charmonium control samples.
(a) $B^+ \to J/\psi  (\to e^+e^-) K^+$,
(b) $B^+ \to J/\psi  (\to \mu^+\mu^-) K^+$,
(c) $B^0 \to J/\psi  (\to e^+e^-) K^{\ast 0}$,
(d) $B^0 \to J/\psi   (\to \mu^+\mu^-) K^{\ast 0}$,
(e) $B^0\to J/\psi (\to e^+e^-) K_S^0$,
(f) $B^0\to J/\psi (\to \mu^+\mu^-) K_S^0$,
(g) $B^+\to J/\psi (\to e^+e^-) K^{\ast +} (\to K_S^0 \pi^+)$, and
(h) $B^+\to J/\psi (\to \mu^+\mu^-) K^{\ast +} (\to K_S^0 \pi^+)$.
The points with error bars show the on-resonance data, and the solid histograms show the 
prediction of the charmonium Monte Carlo. All of the analysis selection criteria 
have been applied except for the charmonium veto, which is reversed. The normalization is absolute.
}
\end{figure}

As we have indicated, the charmonium decays 
$B\to J/\psi K^{(*)}$ and $B \to \psi(2S) K^{(*)}$
provide powerful
control samples for checking the efficiency of our analysis cuts
with events that have properties very similar to those of the
signal. 
Figure~\ref{fig:jpside} compares
the $\Delta E$ distributions (absolutely normalized) 
of these charmonium samples in 
Monte Carlo vs.~data. 
%All of the event selection criteria are applied, 
%except for the charmonium vetos, which are reversed. 
We find good agreement in both the normalization and the shape.

Continuum background is suppressed using a four-variable Fisher 
discriminant, which is a linear combination of the input
variables with optimized coefficients. 
The variables are the Fox-Wolfram moment; the cosine of the angle between 
the $B$ candidate and the beam axis in the CM frame ($\cos \theta_B$); 
the cosine of the angle between the thrust axis of the candidate $B$ meson
daughter particles and that of the rest of 
the event in the CM frame; and the invariant mass $m_{K\ell}$.
The last variable helps discriminate against background from 
$D$ semileptonic decays, for which $m_{K\ell}$ is below the $D$ mass.  
$B\bar{B}$ combinatorial background is suppressed using a likelihood ratio, 
which combines candidate $B$ and dilepton vertex probabilities; 
the significance of the 
dilepton separation in $z$ (along the beam direction); 
$\cos\theta_B$; and the missing energy, $E_{\rm miss}$, of the 
event in the CM frame. The variable $E_{\rm miss}$ provides the strongest
discrimination against $B\bar B$ background: events with two semileptonic
$B$ decays often have more unobserved energy (due to neutrinos) 
than signal events. 

For the purpose of optimizing the event selection, a signal box is 
defined in the $\Delta E$ vs.~$m_{\rm ES}$ plane:
\mbox{5.272 $<$ $m_{\rm ES}$ $<$ 5.286 GeV/$c^2$}  
and \mbox{$-0.11 \ (-0.07) < \Delta E < 0.05$ GeV}
for the electron (muon) channels. 
Using normalizations from the \geant3 based Monte Carlo and 
shapes from the fast Monte Carlo, 
we perform extrapolations to estimate 
the number of expected signal events, $S$, and background events, $B$,
in the signal region for a given set of event selection criteria.
We select the criteria that minimize the probability that the backgrounds
will fluctuate to a level at least as large as the predicted signal.
The result of this optimization is very similar to an optimization 
based on maximizing $S^2/(S+B)$.

%  Fit description

\section{Unbinned Maximum Likelihood Fit}

To extract the signal and background yields, we perform an extended unbinned maximum 
likelihood fit in the $\Delta E$ vs.~$m_{\rm ES}$ plane. 
The fit region is defined by \mbox{$m_{\rm ES}$ $>$ 5.2 GeV/$c^2$} and
\mbox{$|\Delta E|$ $<$ 0.25 GeV}. There are three components in the fit:
the signal, the $B\bar{B}$ background, and the continuum $udsc$ background.
The likelihood is given by
\begin{equation}
{\cal L}(n_{\rm sig},n_{B\bar{B}},n_{\rm cont})=
\frac{e^{-(n_{\rm sig} + n_{B\bar{B}} + n_{\rm cont})}}{N!}
\prod_{i=1}^{N} (n_{\rm sig} P^{\rm sig}_i + 
            n_{B \bar{B}}P^{B\bar{B}}_i +  n_{\rm cont}P^{\rm cont}_i),
\label{eqn:like}
\end{equation}
where $N$ is the total number of events in the fit region; $i$ is an index over
the events;
and $P^{\rm sig}$,
$P^{B\bar{B}}$, and $P^{\rm cont}$ are the normalized probability density functions (which depend on
$m_{\rm ES}$ and $\Delta E$) for the signal, $B\bar{B}$, and continuum background components, respectively.
In the fits to the data, the shapes of these three components are fixed 
(as described below), so that the fit is 
used only to obtain the yields of these components.

The signal shape, $P^{\rm sig}$, must describe a number of important properties of
the signal events, in particular, the effects of photon radiation from the leptons.
The distribution in $m_{\rm ES}$ is essentially a Gaussian centered at the $B$-meson
mass, with a small radiative tail extending to lower values of $m_{\rm ES}$. 
The width of this Gaussian, 
roughly 2.5 MeV$/c^2$ in both electron and muon channels,
is determined primarily by the beam-energy spread. The 
$\Delta E$ distributions are much broader, and
the effects of radiation
are more pronounced.
To describe these shapes and their correlations, 
we fit Monte Carlo signal events to a product of functions known  
as ``Crystal Ball shapes''~\cite{bib:crball}, one each for $m_{\rm ES}$ and 
$\Delta E$. Correlations are taken into account by allowing the 
parameters in the function describing $m_{\rm ES}$ to depend
on $\Delta E$. The function describing $\Delta E$ is actually
somewhat more complicated than the Crystal Ball shape in that
a sum of two Gaussians is used, rather than a single Gaussian,
to help describe the large radiative effects. 

\begin{figure}[!tb]
 \begin{center}
   \includegraphics[height=6.0in]{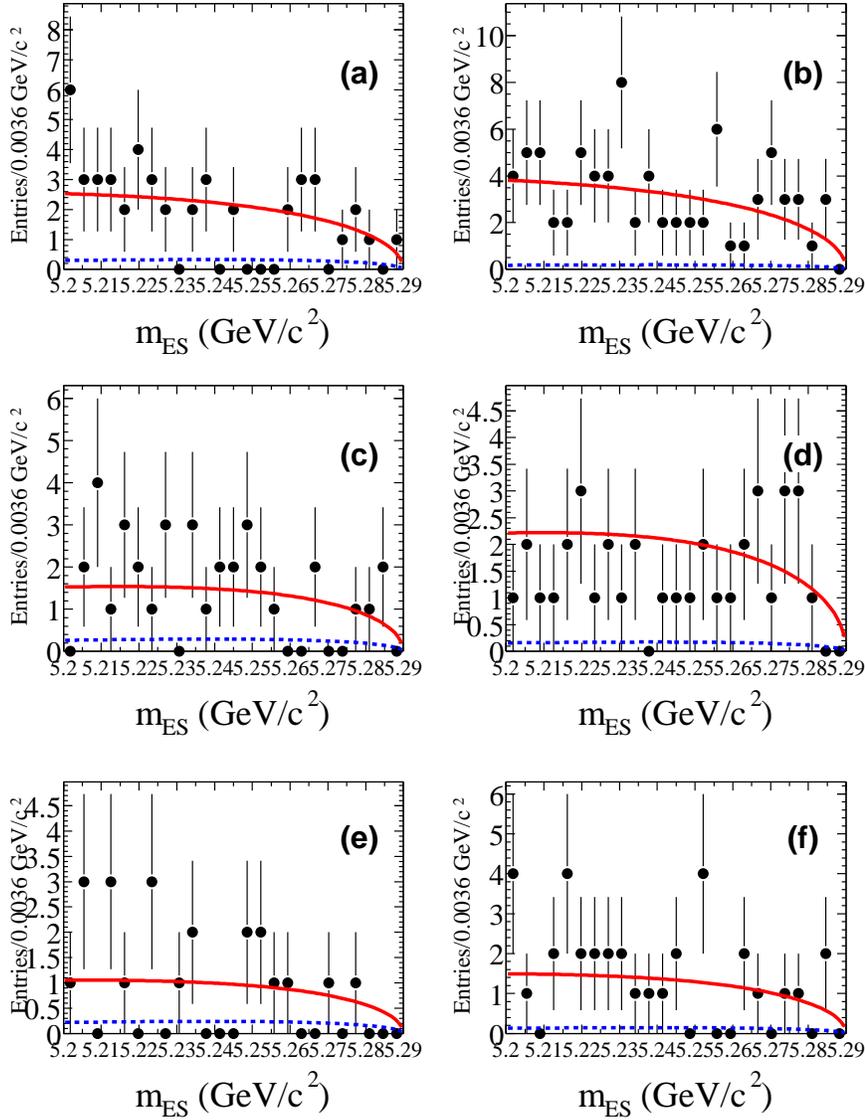}
  \end{center}
  \vspace{-0.5cm}
\caption[Fits to $(K,K^\ast)e\mu$]
{\label{fig:Kemu}
  Fitted $m_{\rm ES}$ distributions in the $(K,K^\ast)e^{\pm}\mu^{\mp}$ 
control sample in data in slices of $\Delta E$:
$\Delta E < -0.11$ GeV for 
(a) \mbox{$Ke^{\pm}\mu^{\mp}$} and (b) \mbox{$K^{\ast0} e^{\pm}\mu^{\mp}$};
$-0.11 < \Delta E < 0.05$ GeV for
(c) \mbox{$Ke^{\pm}\mu^{\mp}$} and (d) \mbox{$K^{\ast0} e^{\pm}\mu^{\mp}$};
and $\Delta E > $ 0.05 GeV for 
(e) \mbox{$Ke^{\pm}\mu^{\mp}$} and (f) \mbox{$K^{\ast0} e^{\pm}\mu^{\mp}$}.
The dashed lines show the continuum component of the fit, and the solid 
lines show the sum
of continuum and $B\bar{B}$ components.
}
\end{figure}

Background events come from two main sources, $\Upsilon(4S)\to B\bar B$
and continuum production of $q\bar q$ pairs. The $B\bar{B}$ background shape, 
$P^{B\bar{B}}$, is parametrized in $\Delta E$ as an exponential of
a second order polynomial in $\Delta E$ multiplied by an Argus function~\cite{bib:argus}
in $m_{\rm ES}$ with the slope parameter taken as a second order polynomial
in $\Delta E$.  
The continuum shape, $P^{\rm cont}$, is taken as the product of
a first order polynomial in $\Delta E$ and an Argus function in $m_{\rm ES}$.
The parameters for the $B\bar B$ and continuum shapes are determined from fitting 
our 200 fb$^{-1}$ sample of fast Monte Carlo events.
The ability of the fast Monte Carlo to predict these background shapes is
checked using the $Ke\mu$ and $K^*e\mu$ control samples in the data
(Figure~\ref{fig:Kemu}). In each final state, the shapes derived from Monte
Carlo describe the data well.

Backgrounds that peak under the signal in $m_{\rm ES}$ and $\Delta E$ 
are not included in the fit, because we have suppressed their
contributions to a level well under one event. 

In the fits we constrain the $B\bar{B}$ and continuum components to be 
greater than or equal to zero. For the signal yield we have a lower cut-off such that 
the total fit function is positive to avoid large negative yields in fits
where there are no events consistent with signal.

%  Results

\section{Results}

Before unblinding the signal region and the near sideband we 
estimated, based on Monte Carlo and off-resonance data, that there 
would be a total of 9.1 background events in 
the signal boxes in the eight modes we analyze. After unblinding, we used the
sideband region to determine the background in the signal
region and found the total background to be 8.4 events.
This is in good agreement with our Monte Carlo and off-resonance  predictions.

\begin{figure}[!tbp]
 \begin{center}
   \includegraphics[height=7.0in]{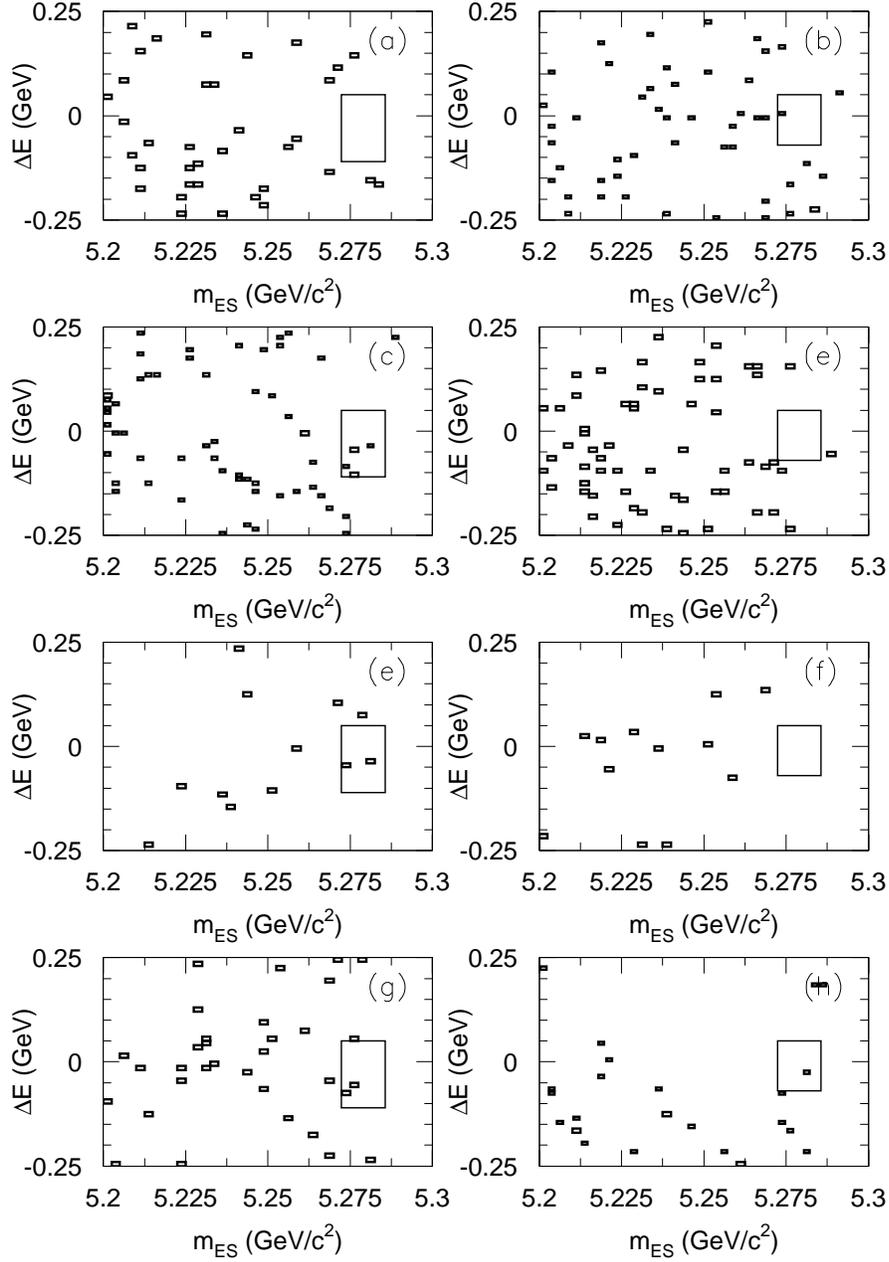}
  \end{center}
 \vspace{-1.0cm}
\caption[Scatterplots.]
{\label{fig:scatterplots}
  Scatterplots of $\Delta E$ vs.~$m_{\rm ES}$ after all analysis cuts in 
(a) $B^+ \to K^+ e^+e^-$,
(b) $B^+ \to K^+ \mu^+\mu^-$,
(c) $B^0 \to K^{\ast 0} e^+e^-$,
(d) $B^0 \to K^{\ast 0}\mu^+\mu^-$,
(e) $B^0 \to K_S^0 e^+e^-$,
(f) $B^0 \to K_S^0 \mu^+\mu^-$,
(g) $B^+ \to K^{\ast +} (\to K_S^0 \pi^+) e^+e^-$, and
(h) $B^+\to K^{\ast +} (\to K_S^0 \pi^+) \mu^+\mu^-$.    
The small rectangle indicates the signal region, which 
is used only for optimizing event selection criteria.
}
\end{figure}

\begin{figure}[!tbp]
 \begin{center}
   \includegraphics[height=7.0in]{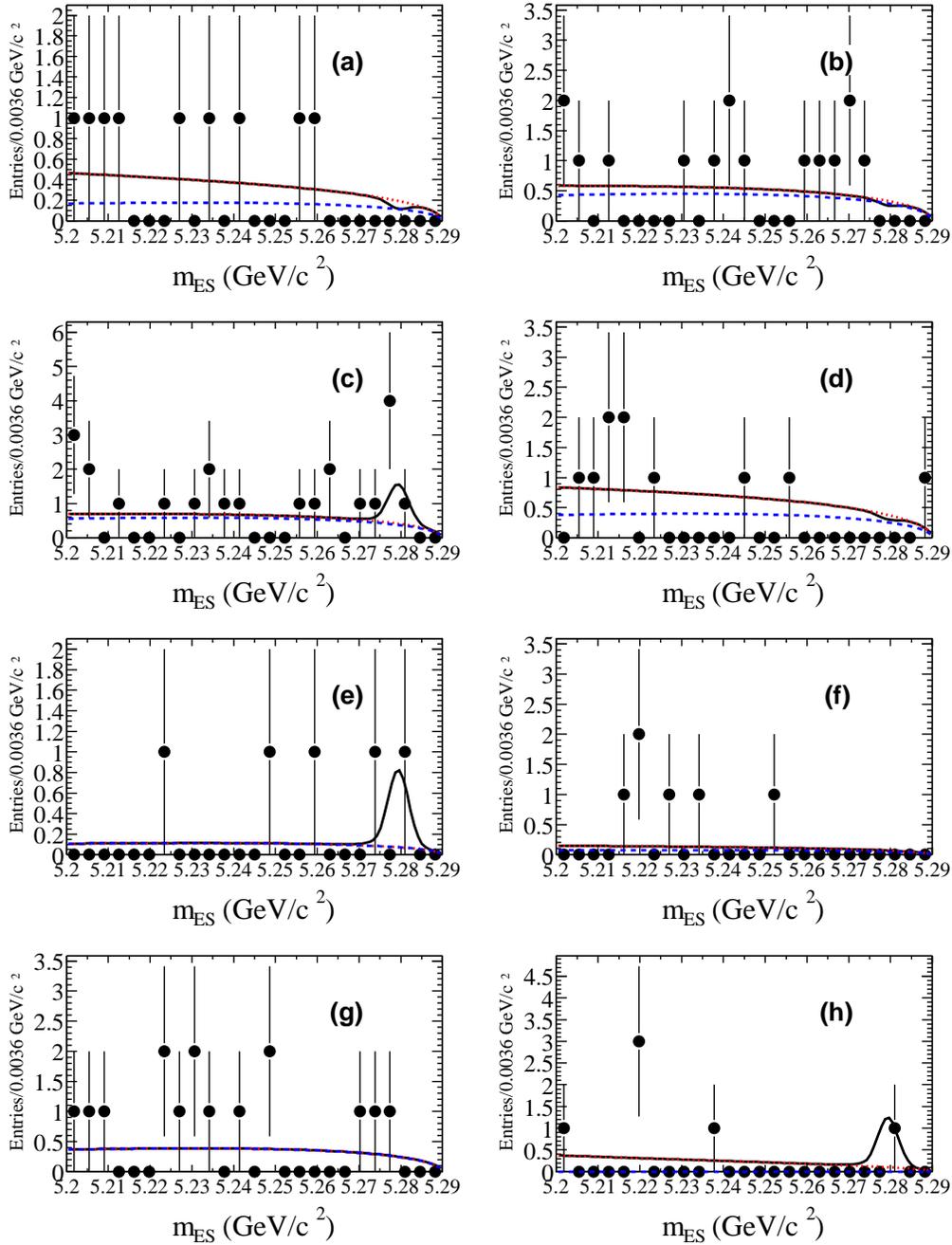}
  \end{center}
  \vspace{-0.5cm}
\caption[Data fit results]
{\label{fig:datafits}
  Fitted $m_{\rm ES}$ distributions in the $\Delta E$ signal region for
(a) $B^+ \to K^+ e^+e^-$,
(b) $B^+ \to K^+ \mu^+\mu^-$,
(c) $B^0 \to K^{\ast 0} e^+e^-$,
(d) $B^0 \to K^{\ast 0}\mu^+\mu^-$,
(e) $B^0 \to K_S^0 e^+e^-$,
(f) $B^0 \to K_S^0 \mu^+\mu^-$,
(g) $B^+ \to K^{\ast +} (\to K_S^0 \pi^+) e^+e^-$, and
(h) $B^+\to K^{\ast +} (\to K_S^0 \pi^+) \mu^+\mu^-$.  
The dashed lines show the continuum component of the fits, 
the dotted lines show the sum
of continuum and $B\bar{B}$ components, and the solid lines show the
sum of background and signal components of the fit.
}
\end{figure}

Figure~\ref{fig:scatterplots} shows the results of unblinding the signal and the near sideband
regions.  We emphasize that the determination of the signal yields is based on the fit, not on
counting the number of events in the signal box.
The fit results are shown in Fig.~\ref{fig:datafits} and are summarized in
Table~\ref{tab:result}. 
By generating multiple toy Monte Carlo samples based on the fit probability
density function (see Eq.~\ref{eqn:like}), we determine the signal
yield such that 90\% of the samples give a number of signal events
larger than that observed in the data.
For the modes where we observed a negative yield, we place the
90\% confidence level limit assuming zero events were observed.
Table~\ref{tab:result} lists an equivalent background yield
determined as the square of the error
on the signal yield in toy Monte Carlo simulations in which only the
background components were included.
Table~\ref{tab:result} also includes the results
for the lepton-number-violating decays $B\to K^{(*)}e\mu$, where the
signal efficiencies were determined from a phase space Monte Carlo simulation
of these decays.

\begin{table*}
\begin{center}
\begin{tabular}{lcccccccc}
 \hline\hline
\multicolumn{1}{c}{Mode}                & Signal  & 90\% C.L. &  Equiv. & $\epsilon$  &  $(\Delta {\mathcal B}/{\mathcal B})_{\epsilon}$  & $(\Delta {\mathcal B}/{\mathcal B})_{\rm fit}$  & ${\mathcal B}/10^{-6}$  & ${\mathcal B}/10^{-6}$ \\
                                        &  yield      & yield     &  bkg. &  (\%)       &    (\%)            & (\%)              &                 &   90\% C.L. \\
 \hline 
$B^+\to K^+e^+e^-$                      &  -0.2  &  3.0      &   0.6  & 17.5        &   $\pm8.6$              &  $\pm10.6$       & -0.1          &  0.9 \\    
$B^+\to K^+\mu^+\mu^-$                  &  -0.2  &  2.8      &   0.4  & 10.5        &   $\pm8.6$              &  $\pm10.6$       & -0.1          &  1.3  \\
$B^0\to K^{*0}e^+e^-$                   &   2.5  &  6.7      &   1.8  & 10.2        &   $\pm10.5$             &  $\pm10.6$       & 1.6            &  5.0  \\
$B^0\to K^{*0}\mu^+\mu^-$               &  -0.3  &  3.6      &   1.1  & 8.0         &   $\pm10.8$             &  $\pm10.6$       & -0.2           &  3.6 \\
$B^0\to K^0 e^+e^-$                   &   1.3  &  5.0      &   0.3  & 15.7        &   $\pm9.3$              &  $\pm10.6$       &  1.1           &  4.7  \\
$B^0\to K^0 \mu^+\mu^-$               &  0.0 &  2.9      &   0.1  & 9.6         &   $\pm11.4$             &  $\pm 10.6$       & 0.0          &  4.5 \\
$B^+\to K^{*+}e^+e^-$           &   0.1 &  3.8      &   0.9  & 8.5         &   $\pm11.4$             &  $\pm10.6$       & 0.1            &  10.0 \\
$B^+\to K^{*+}\mu^+\mu^-$       &   1.0  &  4.3      &   0.5  & 5.8         &   $\pm12.0$             &  $\pm10.6$       & 3.3            &  17.5 \\
%$B^+\to K^{*+}_{\pi^0}e^+e^-$           &   ?   &  ?        &   ?  & ?           &   ?                &  ?          & ?               &  ? \\
%$B^+\to K^{*+}_{\pi^0}\mu^+\mu^-$       &   ?   &  ?        &   ?  & ?           &   ?                &  ?          & ?               &  ? \\
\hline
$B^+\to K^+e^{\pm}\mu^{\mp}$            &  -0.6  &  3.3      &   1.4  & 16.8        &   $\pm8.6$              &  $\pm10.6$       & -0.2           &  1.0 \\
$B^0\to K^{*0}e^{\pm}\mu^{\mp}$         &  0.6   &  4.5      &   2.9  & 11.9        &   $\pm10.6$             &  $\pm10.6$       &  0.3           &  2.7 \\
$B^0\to K^0e^{\pm}\mu^{\mp}$            &  0.8   &  3.3      &   0.7  & 14.6        &   $\pm10.3$             &  $\pm10.6$       &  0.7           &  3.3 \\
$B^+\to K^{*+}e^{\pm}\mu^{\mp}$   & -0.4   &  3.7      &   0.7  & 9.3         &   $\pm11.7$             &  $\pm10.6$       & -0.8           &  8.7 \\
%$B^+\to K^{*+}_{\pi^0}e^{\pm}\mu^{\mp}$ &   ?   &  ?        &   ?  & ?           &   ?                &  10.6       & ?               &  ? \\
% \hline
%$B\to K\ell^+\ell^-$                    &        &           &      &             &                    &            & 0.04             &  0.6 \\
%$B\to K^{*}\ell^+\ell^-$                &        &           &      &             &                    &            & 0.7              &  2.5 \\
 \hline \hline
\end{tabular}
\end{center}
\caption{
\label{tab:result}
Results from the fits to $B\to K^{(\ast)} \ell^+\ell^-$ modes.
The yield refers to the number of signal events given by the fit. The next column
gives the 90\% C.L. upper limit signal yield. The contribution
of the background to the error on the signal yield, expressed as an equivalent background yield (see text),
is reported in the next column. 
$\epsilon$ is the signal efficiency, defined as the efficiency for the 
signal events to be in the fit region and does not include the branching 
fractions of secondary decays.
The systematic error on the efficiency, $(\Delta {\mathcal B}/{\mathcal B})_{\epsilon}$, 
and on the signal yield from the fit, $(\Delta {\mathcal B}/{\mathcal B})_{\rm fit}$, are listed.
The last two columns give the central values of the branching 
fractions (${\mathcal B}$) and the preliminary 
upper limits on the individual modes at 90\% C.L.
}
\end{table*}    
%

%  Systematics section

There are two main categories of systematic uncertainties that affect
the limits: uncertainties from the fit on the extracted number of signal events and 
uncertainties on the signal efficiency. Sources 
of systematic uncertainties are summarized
in Table~\ref{tab:systematic}.

\begin{table}[tbp]
  \begin{center}
    \begin{tabular}{lc}
    \hline\hline
  Source of systematic uncertainty    &  $\Delta {\cal B}/{\cal B}\ (\%)$  \\  \hline
 Lepton tracking efficiency           &  $\pm 1.2$ \\
 Pion from $K^*$ tracking efficiency  &  $\pm 2.0$ \\
 Tracking efficiency for other tracks &  $\pm 1.3$  \\
 Electron identification              &  $\pm 2.7$ \\
 Muon identification                  &  $\pm 2.0$  \\
 Kaon and pion identification         &  $\pm 2.0$ \\
 Monte Carlo statistics for signal    &  $\pm (3.0\ {\rm to}\ 5.0)$ \\
 Continuum suppression cut            &  $\pm 2.0$  \\
 $B\bar B$ suppression cut            &  $\pm 3.0$  \\
 $K_S$ selection                      &  $\pm 4.0$  \\
 Modeling of signal kinematic distributions   &  $\pm 5.0$ \\
 Number of $B\bar B$ pairs            &  $\pm 1.6$ \\
 Background shapes                    &  $\pm 7.0$ \\
 Signal shapes                        &  $\pm 8.0$ \\
    \hline \hline
    \end{tabular}
  \end{center}
  \caption[Systematic uncertainty]
    {\label{tab:systematic}
       Sources of systematic uncertainties on the limits.
   }
\end{table}

%Additionally, there is a $\pm 1.6\%$ uncertainty on the number of produced
%$B$ mesons in the data sample.
%
%The systematic errors on the signal efficiency include the uncertainty on the
%tracking efficiency ($\pm 1.2\%$ for leptons and $\pm 1.3\%$ for all other tracks,
%except the relatively slow pions from $K^\ast$ decays, where it is $\pm 2\%$);
%the uncertainty on the particle identification efficiency 
%($\pm 2.7\%$ for electrons,  $\pm 2.0\%$ 
%for muons, and $\pm 2\%$ for charged kaons and pions); the statistical
%uncertainty due to the size 
%of the signal Monte Carlo samples (about $\pm 3\%$ for $B\to K\ell^+\ell^-$
%and about $\pm 5\%$ for $B\to K^{\ast0}\ell^+\ell^-$); the uncertainty on the
%modeling of the efficiency for continuum suppression cuts ($\pm 2\%$), 
%on the $B\bar{B}$ suppression cuts
%($\pm 3\%$), and on the $K_S^0$ finding criteria for the modes with a $K_S^0$ ($\pm 4\%$); 
%and finally the systematic error on the efficiency arising from the  
%theoretical uncertainty in the kinematic distributions ($\pm 5\%$).  
%We have considered the models
%of Refs.~\cite{bib:Colangelo1999} and~\cite{bib:Melikhov1998}, and found 
%that the signal efficiency varies from 2 to 5\%,  depending on the mode.

The sources of systematic uncertainty affecting the extraction of the signal yield are
the errors associated with the choice of the background and
signal parametrization in the fit.  To evaluate the systematic uncertainty on the background shape, 
we took the difference between the nominal result for the 90\% C.L.~event yield and the  
largest deviation found if one allows the background shape to be given either by a
pure $B\bar B$ or a pure continuum shape.
We assign an error of $\pm 7\%$ to the uncertainty in the knowledge of the background shape.
The systematic error on the signal parametrization is established by 
varying the signal shape using the parameters obtained from fitting the charmonium control sample,
and is of the order $\pm 8\%$.

In setting the upper limit, the systematic errors on the efficiency,
$(\Delta {\cal B}/{\cal B})_{\epsilon}$,
and on the signal yield from the fit, $(\Delta {\cal B}/{\cal B})_{\rm fit}$,
are added in quadrature, and the limit is increased by the corresponding
factor.

%  Conclusion section

\section{Combined Results and Conclusions}

We average the electron and muon channels to determine the average
branching fractions 
${\cal B}(B\to K\ell^+\ell^-)$ and ${\cal B}(B\to K^*\ell^+\ell^-)$.
The modes are combined using the likelihood fit where a combined
likelihood is formed as a product of the likelihoods of the individual
modes. The fit extracts a combined signal event yield. For the 
averaging of modes with $K^*$ mesons
the ratio of branching fractions 
${\cal B}(B\to K^*e^+e^-)/{\cal B}(B\to K^*\mu^+\mu^-)$ from the model
of Ali {\it et al.} (see Table~\ref{tab:predictions}) is used to
weight the yield in the muon mode relative to the electron mode. The
extracted yield corresponds to the electron mode.
The combined fits give
\begin{eqnarray}
{\mathcal B} (B\to K\ell^+\ell^-) & = &(0.0\pm0.3({\rm stat.}))\times 10^{-6}   \nonumber\\
{\mathcal B} (B\to K^*\ell^+\ell^-) & = &(0.7\pm 1.1({\rm stat.}))\times 10^{-6}  \nonumber.
\end{eqnarray}
As there is no evidence for a signal we evaluate the upper limits
on these combined modes and obtain the preliminary results
\begin{eqnarray}
{\mathcal B} (B\to K\ell^+\ell^-) & < &0.6\times 10^{-6}  \  {\rm at}\ 90\%  \ {\rm C.L.}\ \nonumber\\
{\mathcal B} (B\to K^*\ell^+\ell^-) & < &2.5\times 10^{-6} \   {\rm at}\ 90\%  \ {\rm C.L.}\nonumber
\end{eqnarray}
based on the analysis of a sample of $22.7\times 10^6$ $\FourS\to B\bar B$
decays in the \babar\ 1999-2000 data set. 
These limits represent a significant improvement over previous results~\cite{bib:CDF,bib:CLEO,bib:BELLE}
and are at the same level as the Standard Model predictions listed in
Table~\ref{tab:predictions}.

%%%%%%%%%%%%%%%%%%%%%%%%%%%%%%%%%%%%%%%%%%%%%%%%%%%%%%%%%%%%%%%%%%%%%%%%

% Specific acknowledgments for this paper; remove if not needed.
%The authors wish to thank Prof.\ A.\ Cleverguy for his help
%with the theoretical interpretation of these results.

% Standard acknowledgments paragraph; must always be included.
\section{Acknowledgements}
We are grateful for the 
extraordinary contributions of our \pep2\ colleagues in
achieving the excellent luminosity and machine conditions
that have made this work possible.
The collaborating institutions wish to thank 
SLAC for its support and the kind hospitality extended to them. 
This work is supported by the
US Department of Energy
and National Science Foundation, the
Natural Sciences and Engineering Research Council (Canada),
Institute of High Energy Physics (China), the
Commissariat \`a l'Energie Atomique and
Institut National de Physique Nucl\'eaire et de Physique des Particules
(France), the
Bundesministerium f\"ur Bildung und Forschung
(Germany), the
Istituto Nazionale di Fisica Nucleare (Italy),
the Research Council of Norway, the
Ministry of Science and Technology of the Russian Federation, and the
Particle Physics and Astronomy Research Council (United Kingdom). 
Individuals have received support from the Swiss 
National Science Foundation, the A. P. Sloan Foundation, 
the Research Corporation,
and the Alexander von Humboldt Foundation.

%%%%%%%%%%%%%%%%%%%%%%%%%%%%%%%%%%%%%%%%%%%%%%%%%%%%%%%%%%%%%%%%%%%%%%%%

\end{document}